\documentclass{PoS}

\title{Calibration of the INTEGRAL SPI Anti
Coincidence Shield with Gamma Ray Bursts observations}

\ShortTitle{Calibration of INTEGRAL SPI/ACS with GRBs}

\author{D.~Vigan\`o$^{a,b}$ and S.~Mereghetti$^b$\\
        \llap{$^a$}Dipartimento di Fisica, Universit\`a degli Studi di Milano,\\
	via G.~Celoria, 16, I 20133 Milano, Italy\\
	\llap{$^b$}INAF-IASF Milano,\\
	via E.~Bassini, 15, I 20133 Milano, Italy\\
	E-mail: \email{vigano@iasf-milano.inaf.it}, \email{sandro@iasf-milano.inaf.it}}

\abstract{The Anti Coincidence Shield (ACS) of the INTEGRAL SPI instrument provides an excellent sensitivity for the detection of Gamma Ray Bursts (GRBs) above $\sim$75 keV, but no directional and energy information is available. We studied the ACS response by using GRBs with known localizations and good spectral information derived by other satellites. We derived a count rate to flux conversion factor for different energy ranges and studied its dependence on the GRB direction and spectral hardness. For a typical GRB spectrum, we found that 1 ACS count corresponds on average to $\sim 10^{-10}\mbox{ erg/cm}^2$ in the 75 keV-1 MeV range, for directions orthogonal to the satellite pointing axis. This is broadly consistent with the ACS effective area derived from the Monte Carlo simulations, but there is some indication that the latter slightly overestimates the ACS sensitivity, especially for directions close to the instrument axis.}

\FullConference{The Extreme sky: Sampling the Universe above 10 keV - extremesky2009,\\
		October 13-17, 2009\\
		Otranto (Lecce) Italy}

\begin{document}

\section{Introduction}
The Anti Coincidence Shield (ACS) of the INTEGRAL \cite{winkler03} SPI instrument \cite{vedrenne03} consists of 91 BGO crystals  covering the lateral and bottom sides of the spectrometer \cite{vonKienlin01}. The ACS provides veto signals for charged particles and gamma rays coming from outside the field of view (FoV). Its effective area, depending on energy and direction, has been computed for a set of directions and energies by means of Monte Carlo simulations \cite{weidenspointner05}, but no dedicated in-flight calibrations were performed. Using Gamma Ray Bursts (GRBs) it is possible to derive at least a rough in-flight calibration, i.e. a conversion factor from the instrument counting rate to physical flux units. This requires the knowledge of  spectral and positional data of GRBs seen by at least another satellite besides the ACS.

\section{Sample selection and properties}
Based mainly on the information reported in the \emph{Gamma Ray Bursts Coordinate Network}\footnote{\texttt{http://gcn.gsfc.nasa.gov/gcn/gcn3\_archive.html}} (GCN) and in the on-line Swift catalogue\footnote{\texttt{http://swift.gsfc.nasa.gov/docs/swift/archive/grb\_table.html/}} we collected for each GRB the occurrence time and coordinates, fluence,  peak flux, duration ($T_{90}$) and best fit spectral parameters. We considered a period of 6.5 years,  from $1^{st}$ January, 2003, few months after the launch of INTEGRAL, to $30^{th}$ June, 2009.
This resulted in a total of 764 GRBs, the majority of which were detected by Swift-BAT, Fermi-GBM and Konus-WIND (hereafter: BAT, GBM, KW). Table \ref{instruments} shows, for each satellite, the period of activity, the total number of GRBs, the number of GRBs for month and the number of  GRBs in common with the ACS.

\begin{table}[ht!]
\begin{center}
\begin{footnotesize}
\begin{tabular}{|l|c|c|c|c|}
\hline
Instrument 	& Period 	   & GRBs tot & GRBs/month & detected by ACS \\
\hline
Swift-BAT 	& 12/2004 - 06/2009 & 440 &  8.1 & 90 \\
Fermi-GBM 	& 08/2008 - 06/2009 & 174 & 17.4 & 80 \\
Konus-WIND	& 01/2003 - 06/2009 & 120 &  1.6 & 79 \\
Suzaku-WAM	& 08/2005 - 06/2009 &  73 &  1.6 & 42 \\
HETE	  	& 01/2003 - 03/2006 &  54 &  1.4 & 17 \\
INTEGRAL-IBIS	& 01/2003 - 06/2009 &  54 &  0.7 &  5 \\
Ulysses		& 01/2003 - 05/2003 &  17 &  3.4 & 13 \\
AGILE		& 05/2007 - 06/2009 &  11 &  0.4 &  4 \\
RHESSI		& 01/2003 - 06/2009 &   6 &  0.1 &  6 \\
\hline
\end{tabular}
\end{footnotesize}
\caption{GRBs reported in the GCN during the indicated period.}
\label{instruments}
\end{center}
\end{table}

Not all these bursts are visible in the light curves collected by the ACS, for several reasons, as indicated in Table \ref{grb_acs_lc}. Furthermore, some of the bursts visible in the ACS have missing fluence or spectral parameters, or very coarse positional information $\sigma_{pos}>15^\circ$. We therefore remain with 196 GRBs, but since a few of them have structured light curves, with several peaks that could be analyzed separately, our final sample consists of 205 events. \bigskip

\begin{table}[ht!]
\begin{center}
\begin{footnotesize}
\begin{tabular}{|l|c|c|}
\hline
Events						& N 	& $\%$ \\
\hline
\textbf{Visible in ACS light curves}	&\textbf{196}	& \textbf{25.6} \\
Visibile, but with poor spectral/positional information	&  18 &  2.4 \\
Uncertain (S/N very low)			&  28	&  3.7 \\
Not visible					& 132	& 17.3 \\
Off (e.g.: INTEGRAL perigee passage)		&  16	&  2.1 \\
SPI/ACS light curve not present in the catalogue& 374	& 48.9 \\
\hline
TOTAL						& 764	& 100.0 \\
\hline
\end{tabular}
\end{footnotesize}
\caption{GRBs in the ACS light curves.}
\label{grb_acs_lc}
\end{center}
\end{table}

\begin{figure}
\centering
\includegraphics[width=60mm,angle=0]{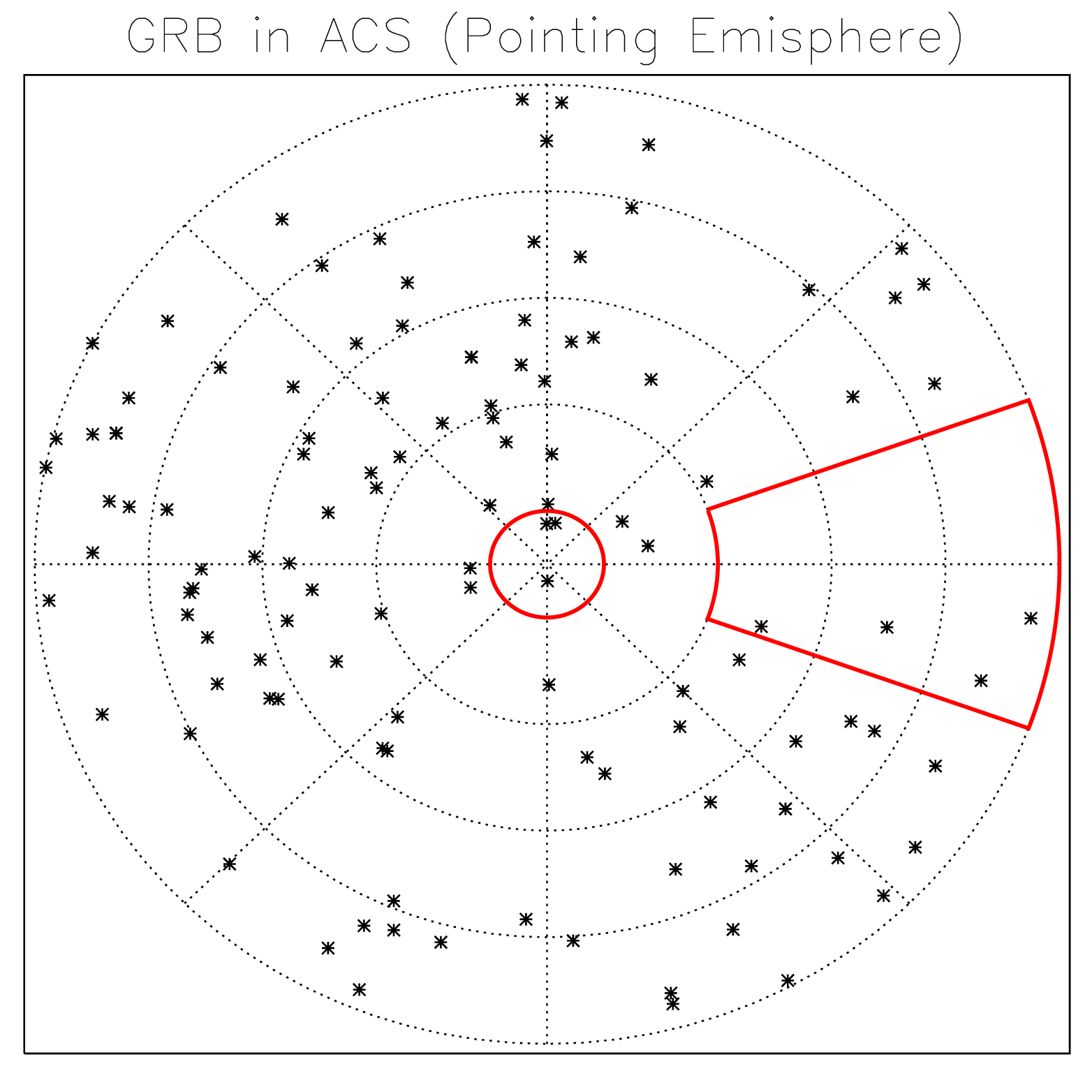}
\includegraphics[width=60mm,angle=0]{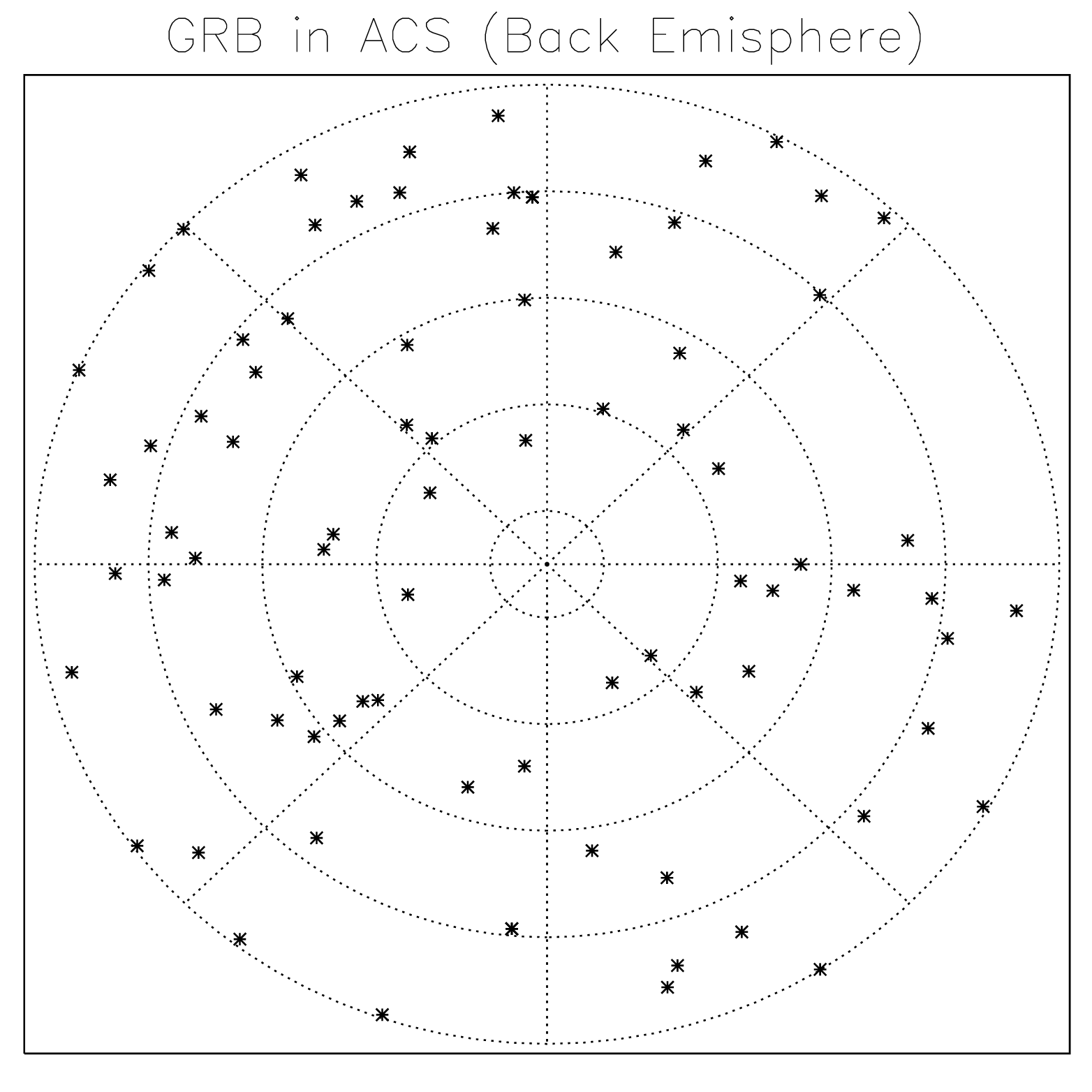}
\caption{Directions of 196 good GRBs in instrumental coordinates. The two plots refer to the front emisphere (centered at the pointing direction) and the back emisphere. The areas delimited by the red thick lines indicate the SPI instrument FoV and the approximate area obscured by the IBIS instrument.}
\label{ACS_GRB_map}
\end{figure}

Fig. \ref{ACS_GRB_map} shows the distribution of our sample in detector coordinates. As expected, more GRBs are detected by the ACS in the directions orthogonal to the satellite pointing direction, except for the side partially shielded by the IBIS instrument.
\par
\begin{figure}
\centering
\includegraphics[height=60mm,angle=0]{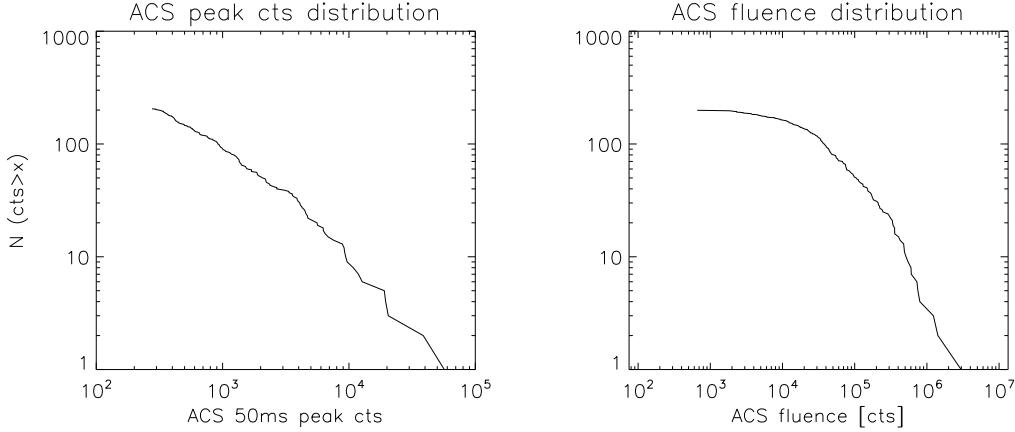}
\caption{Integral distributions of background-subtracted peak fluxes [counts/bin] and fluences [counts] of our sample.}
\label{sample_prop}
\end{figure}

The instrumental background in the ACS is produced mainly by particle interactions and its value in the considered period ranges from 3500 to 7500 counts/bin (1 bin=50 ms), with typical standard deviation of 100 counts/bin. Fig. \ref{sample_prop} shows the integral distributions of background-subtracted peak fluxes $PC_{ACS}$ and fluences $N_{ACS}$, in units of counts/bin and counts, respectively. The values of the peak fluxes are in the range 275-57000 counts/bin, while the values of the fluences extend between 660 and $3\times 10^6$ counts.

\subsection{Spectral sample}
The instruments best suited to derive the ACS calibration are KW and GBM, thanks to the large number of GRBs in common with the ACS and to their energy ranges (15 keV-10\,MeV and 10\,keV-30\,MeV) which overlap quite well the ACS one. After rejecting a few events, due to incomplete ACS light curves or lack of reported errors in some of their spectral parameters, we could define a \emph{spectral sample} of 133 events: 62 seen by GBM, 71 by KW. The reported spectral results were in most cases based on the Band  \cite{band93} (BF) or  Cut-off Power Law (CPL) models. If we define $\alpha$ the low-energy photon index (present in CPL and BF), $\beta$ the high-energy photon index (BF), $E_0$ the cut-off energy (CPL) or break energy (BF), we obtain the following mean values for the spectral sample:

\begin{equation}
\bar{\alpha}=-0.86\pm 0.30, \quad \bar{\beta}=-2.31\pm 0.30, \quad \bar{E_0}=448\pm 298\,\mbox{keV}
\label{parametri_sample}
\end{equation}

\section{Derivation of ACS count rate to flux conversion}
We found that the ACS conversion factors derived using the peak fluxes have a wider dispersion than those derived using the fluences. This is due to the fact that peak fluxes have larger statistical errors and less constrained spectral parameters. Furthermore, the reported values often refer to different time integration intervals, thus introducing another source of uncertainty in the ACS comparison. We therefore based our analysis on the GRB fluences, computing for each burst the following quantity
\begin{equation}
k\equiv\frac{f_{ACS}[10^{-7}\mbox{erg/cm}^2]}{N_{ACS}[1000\,\mbox{counts}]}
\label{def_k}
\end{equation}
\noindent where $N_{ACS}$ is the measured fluence in ACS counts and $f_{ACS}$ is the fluence in physical units obtained by converting the KW or GBM results to the ACS energy range. In this conversion we took into account the 90\% c.l. errors on all the parameters in order to estimate the error on $k$. We defined the conversion factor as $\bar{k}$, i.e. the weighted average of $k$. The conversion factor depends on the energy range used for $f_{ACS}$. We have assumed $E_{min}$=75\,keV, and different values for $E_{max}$ as indicated in the first column of Table \ref{tab_calib_zone}.

\begin{footnotesize}
\begin{table}
\centering
\begin{tabular}{|c|c c|c c|c c|c c|}
 \hline
 range [MeV] & $\bar{k}$ & $\tilde{\chi}^2$ &$\bar{k}_{top}$ & $\tilde{\chi}^2$ & $\bar{k}_{cnt}$ & $\tilde{\chi}^2$ & $\bar{k}_{bot}$ & $\tilde{\chi}^2$\\
 \hline
 $ 0.075-1$ & $0.98\pm0.01$ & 57 & $2.14 \pm 0.07$ & 17  & $0.82 \pm 0.01$ & 317 & $2.48\pm0.03$ & 17 \\
 $ 0.075-2$ & $1.37\pm0.02$ & 23 & $2.54 \pm 0.08$ & 15  & $1.18 \pm 0.02$ &  89 & $3.94\pm0.09$ &  6 \\
 $ 0.075-5$ & $1.77\pm0.02$ & 12 & $2.62 \pm 0.10$ &  7  & $1.61 \pm 0.02$ &  32 & $3.88\pm0.11$ &  9 \\
 $0.075-10$ & $1.90\pm0.03$ &  9 & $2.58 \pm 0.11$ &  6  & $1.73 \pm 0.03$ &  18 & $3.85\pm0.12$ &  9 \\
 \hline
\end{tabular}
\caption{Conversion factors derived for the whole spectral sample (133 events), 27 top zone events ($\theta<45^\circ$), 78 central zone events ($45^\circ<\theta<120^\circ$), 22 bottom zone events ($\theta>120^\circ$).}
\label{tab_calib_zone}
\end{table}
\end{footnotesize}

\begin{figure}
\centering
\includegraphics[height=80mm,angle=0]{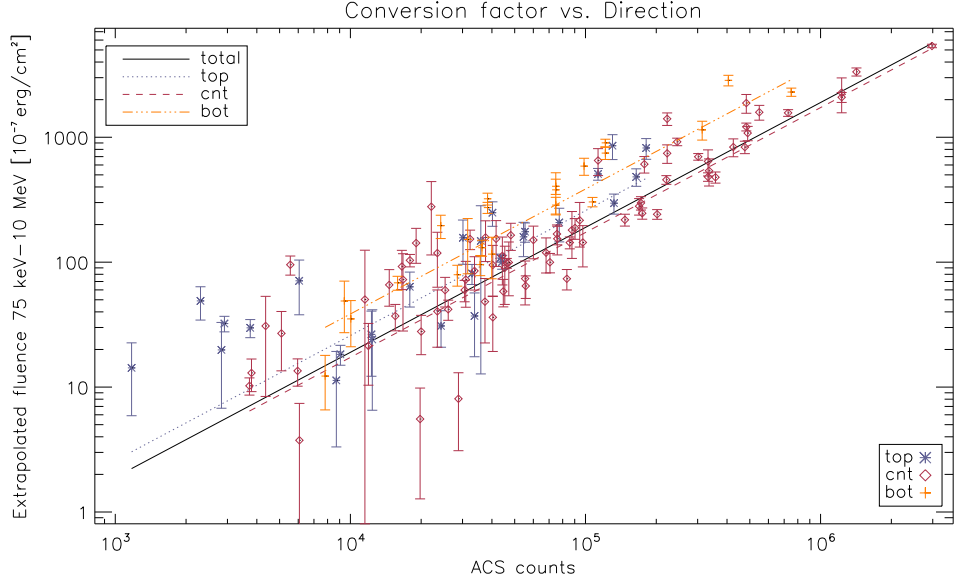}
\caption{Correlation between counts and fluence for GRBs in the total sample and in the three zones defined by the angles $(\theta_1,\theta_2)=(45^\circ,120^\circ)$, with $E_{max}=10$ MeV.}
\label{k10_zone}
\end{figure}

The large dispersion of $k$ around the mean value (see Fig. \ref{k10_zone} and values of $\tilde{\chi}^2$ in Table \ref{tab_calib_zone}) can be ascribed  to the directional and spectral variety of the sample. To investigate this effect, we divided our spectral sample in three subsamples: top zone events ($\theta<45^\circ$), central zone events ($45^\circ<\theta<120^\circ$, excluding IBIS obscured zone), bottom zone events ($\theta>120^\circ$), where $\theta$ is the angle from the pointing direction. Table \ref{tab_calib_zone} gives the resulting values of $\bar{k}$. As expected, the values of $\bar{k}_{cnt}$ are smaller than $\bar{k}_{bot}$ and $\bar{k}_{top}$, because the ACS sensitivity is larger for orthogonal directions. However, the large dispersion of $k$ values also for the three individual subsamples indicates that they also depend significantly on the spectral variety.
Therefore, we studied the dependence of $k$ on the GRB spectra, that we characterized by the hardness ratio defined as follows. We chose a threshold energy $E_T$=500\,keV, near  $\bar{E_0}$ shown in Eq. \ref{parametri_sample}. We defined for each burst the soft and hard parts of the extrapolated fluence $S\equiv f_{75\,keV-E_T}$, $H\equiv f_{E_T-E_{max}}$ and the hardness ratio $HR=(H-S)/(H+S)$. For every choice of $E_{max}$ we found that the dispersion of $k$ rises with increasing $HR$. This is probably due to the larger uncertainties in the extrapolated fluences at higher energy, affecting especially the hardest bursts. Anyway a clear trend between $HR$ and $\bar{k}$ is not visible. A different choice of $E_T$ does not yield better correlations.

\section{Discussion}
We performed a simple comparison of our results with the conversion factors estimated by means of Monte Carlo simulation \cite{weidenspointner05}. Based on the values of the effective area between $80$ and $500$ keV computed for a set of $19\times 6$ directions ($\theta$, $\phi$), we derived the expected conversion factor for an input photon spectrum $N(E)$:

\begin{equation}
 k_{sim}(E_{max},\theta,\phi)=\frac{\int^{E_{max}}_{E_{min}}\, E\, N(E)\,\mbox{d}E}{\int^{E_{max}}_{E_{min}}\, A_{eff}(E,\theta,\phi)\,N(E)\,\mbox{d}E} \frac{[10^{-7}\mbox{erg/cm}^2]}{[1000\,\mbox{counts}]}
\end{equation}
\noindent $A_{eff}$ shows almost no dependence on $\phi$, except for the direction obscured by IBIS. So we evaluated $k_{sim}(\theta)$ for the mean spectrum described in Eq. \ref{parametri_sample}. 
$k_{sim}$ ranges between $\sim 1.0$ and $\sim 1.6$ for $E_{max}$=10\,MeV and between $\sim 0.7$ and $\sim 1.0$ for $E_{max}$=1\,MeV. These values are smaller than those derived from our analysis of GRBs (see Table \ref{tab_calib_zone}), especially for the top and bottom zone.
\par
We repeated this analysis with two different spectra: the softest and the hardest resulting from spectral parameters within $1\sigma$ of the average values. For any fixed value of $\theta$ and $E_{max}$ we can see a difference between the resulting values of $k_{sim}$ consistent with the observed experimental dispersion.

\section{Conclusions}
Using well localized GRBs, we have derived an in-flight ACS counting rate to physical units conversion factor. Despite the dispersion due to the positional and spectral variety, we could obtain a correlation between the response and the direction of GRB. For a typical GRB spectrum and directions orthogonal to the SPI axis, the average conversion factor between counts and fluence in the range 75 keV-1 MeV is
\begin{equation}
 1\mbox{ ACS count}\sim 10^{-10}\mbox{ erg/cm}^2
\end{equation}
\noindent while the conversion factor in the non-orthogonal directions is a factor $\sim 2-3$ larger.
\par
This project could be developed in the near future, when a fastly increasing amount of Fermi/\\GBM data will be available. A larger GRB sample can improve the accuracy of the directional and energy dependence of the conversion factor. In particular, the selection of a more spectrally uniform subsample is suggested in order to partially get rid of the data dispersion.
\par
\bigskip
\noindent \textbf{Acknowledgements}. This work has been partially supported by ASI contract I/008/07/0.

\end{document}